\documentclass{PoS}

\newcommand{\braket}[1]{\langle #1 \rangle}
\newcommand{\Tprod}[1]{{\mathrm T}\lbrack #1 \rbrack}
\newcommand{\grtsim}{\mbox{\raisebox{-3pt}{$\stackrel{>}{\sim}$}}}

\title{Electroweak non-resonant corrections to top pair production close to threshold}

\ShortTitle{Electroweak non-resonant corrections to top pair production close to threshold}

\author{M.~Beneke$^a$,  B.~Jantzen$^a$ and \speaker{P.~Ruiz-Femen\'ia}~$^{ab}$
\thanks{Preprint numbers: TTK-10-53, SFB/CPP-10-124. This work is supported by the DFG Sonder\-forschungsbereich/Transregio~9
``Computergest\"utzte Theoretische Teilchenphysik''.}
       \\
        \llap{$^a$}Institut f\"ur Theoretische Teilchenphysik und Kosmologie, RWTH Aachen University\\
                          D-52056 Aachen, Germany\\
\llap{$^b$}Fakult\"at f\"ur Physik, Universit\"at Wien, 1090 Wien, Austria\\
        E-mail: \email{ruiz@physik.rwth-aachen.de}}

\abstract{The production of  $W^+W^- b \bar{b}$ from $e^+e^-$ collisions at energies close to the $t\bar{t}$ threshold is dominated by the resonant process with a nearly on-shell $t\bar{t}$ intermediate state. The
 $W b$ pairs in the  final state can also be reached through the decay of off-shell tops or through background processes containing no or only single top quarks. 
This non-resonant production starts to contribute at NLO to the $W^+W^- b \bar{b}$ total cross section in the non-relativistic power-counting  $v ~\sim \alpha_s \sim \sqrt{\alpha_{EW}}$. The NLO non-resonant corrections
 presented in this talk represent the non-trivial NLO electroweak corrections to the
 $e^+e^- \to W^+W^- b \bar{b}$ cross section in the top anti-top resonance region. In contrast to the QCD corrections which have been calculated (almost) up to NNNLO, the parametrically larger NLO electroweak contributions have not been completely known so far, but are mandatory for the required accuracy at a future linear collider.  We consider the total cross section of the $e^+e^- \to W^+W^- b \bar{b}$  process and additionally implement cuts on the invariant masses of the $W^+ b$ and $W^-  \bar{b}$ pairs.}

\FullConference{35th International Conference of High Energy Physics - ICHEP2010,\\
		July 22-28, 2010\\
		Paris France}

\begin{document}

\section{Introduction}

The top-quark mass is currently known from direct production at the 
Fermilab Tevatron (and soon at the Large Hadron Collider) with 
a precision $\grtsim 1$~GeV. From a 
threshold scan of the $e^+ e^-\to t\bar t$ cross section 
at the planned International Linear Collider (ILC), however, 
an order of magnitude improvement in the precision can be achieved 
experimentally~\cite{Martinez:2002st}. Aside from determining a 
fundamental parameter of the Standard Model, accurate top-mass
measurements constrain the quantum fluctuations from 
non-standard interactions in electroweak precision measurements. 
Other characteristics of the top quark 
such as its width and Yukawa coupling provide information 
about its coupling to other particles and the mechanism of 
electroweak symmetry breaking. For these reasons top-quark pair 
production near threshold in $e^+ e^-$ annihilation has been 
thoroughly investigated following the 
non-relativistic QCD (NRQCD) approach,
which treats the leading colour-Coulomb force exactly to all orders 
in perturbation theory. In this framework, where the strong 
coupling $\alpha_s$ is of the same order as $v$,
the small relative velocity of the top and anti-top, 
most QCD corrections to the total cross section have been calculated
up to NNNLO (see the summary~\cite{Beneke:2008ec}),
and next-to-next-to-leading logarithms of $v$ have been
partially summed~\cite{Hoang:2001mm}.

Here we focus on the subleading electroweak corrections, 
which have received much less attention.
The top quark is unstable with a 
significant width $\Gamma_t$ of about $1.5\,$GeV due to the electroweak 
interaction. The width is essential in threshold production, since 
it prevents the top and anti-top from forming a bound 
state and causes a broad resonance structure 
in the energy dependence of the cross section on top of the 
increase due to the opening-up of the two-particle phase space. 
Once the top width is included, due to top decay, the physical 
final state is $W^+ W^- b\bar b$ -- at least 
if we neglect the decay of top into strange and down quarks, as 
justified by $V_{tb}\approx 1$, and consider $W$ bosons as 
stable. The $W^+W^- b \bar{b}$ final state can also be produced 
non-resonantly, {\it i.e.} through processes which do not involve  a nearly on-shell
$t\bar{t}$ pair.   
The latter effects are not included in the standard
non-relativistic treatment. Adopting a counting scheme where 
$\alpha_{\rm EW}\sim \alpha_s^2$, we find that the leading 
non-resonant and off-shell effects are NLO for 
the total cross section, since there is an additional power 
of $\alpha_{\rm EW}$ but no phase-space suppression, hence 
the relative correction is $\alpha_{\rm EW}/v\sim \alpha_s$.
Purely resonant electroweak effects, on the other hand, yield
 NNLO corrections~\cite{Hoang:2004tg}.

In this talk we present the calculation of the non-resonant NLO electroweak
contributions to the $e^+ e^-\to W^+ W^- b\bar b$ process 
in the $t\bar t$ resonance 
region, for the total cross section as well as including 
invariant-mass cuts on the $Wb$ pairs.
The calculation is performed 
with unstable-particle effective field 
theory~\cite{Beneke:2003xh}, which provides the  
framework for consistently including resonant and non-resonant 
effects while maintaining an expansion in the small parameters 
of the problem.

\section{Unstable-particle effective theory for pair production near threshold}

%
\begin{figure}[t]
\vskip -.5cm
\includegraphics[width=0.4\textwidth]{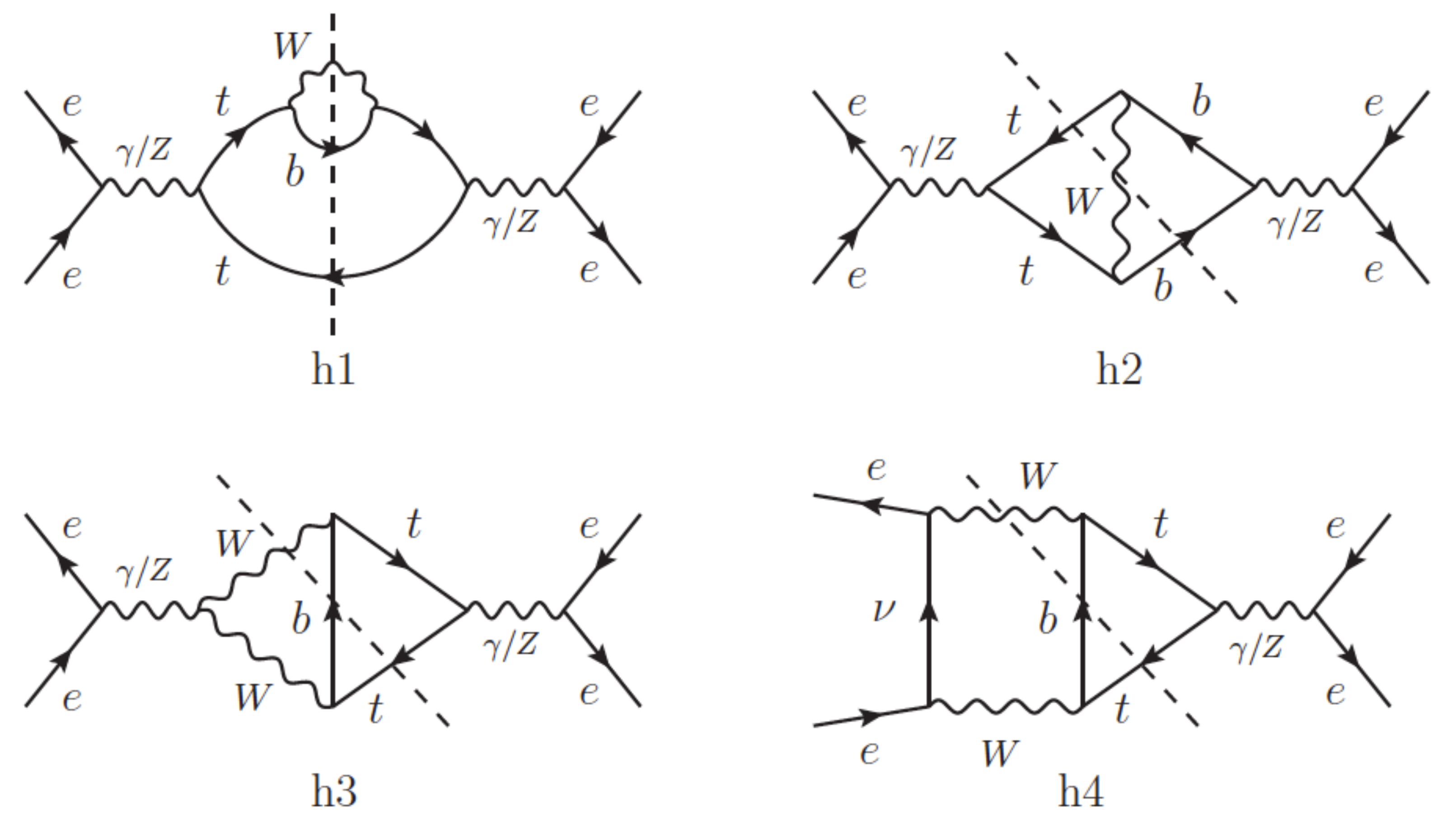}\\
{\vskip -3.9cm
\hskip 6.5cm
\includegraphics[width=0.55\textwidth]{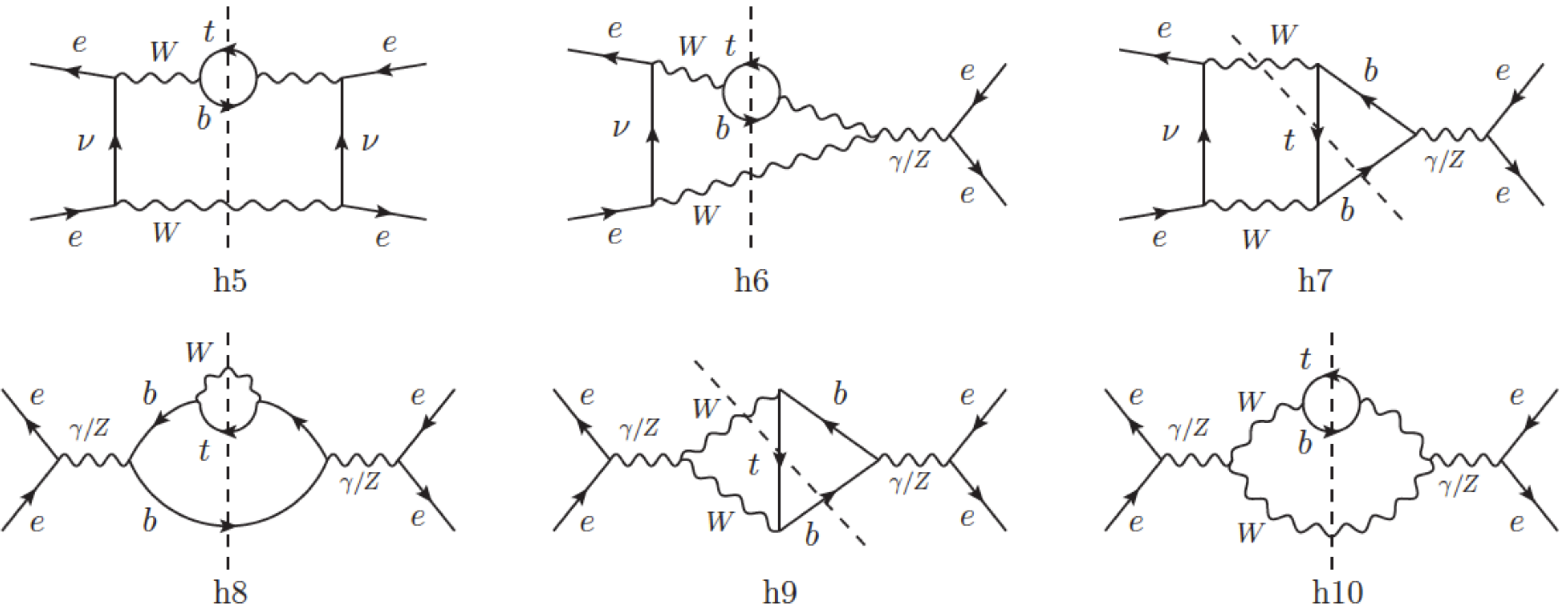}}
\vskip -.2cm
\caption{Two-loop forward-scattering amplitude diagrams with 
$\bar{t} b W^+$ cuts. $t \bar{b} W^-$ cuts and symmetric diagrams 
are not shown. The contribution to the $W^+W^-b\bar{b}$ cross section from diagrams 
$h_1$--$h_{10}$ can be interpreted as the $\bar{b} W^-$ pair
originating from a nearly on-shell anti-top decay, while the $b W^+$ pair
is produced non-resonantly, either from a highly virtual top 
(diagrams $h_1$--$h_4$), or without an intermediate top 
($h_5$--$h_{10}$).}
\label{fig1}
\vspace*{-0.2cm}
\end{figure}
The cross section for the 
$e^+ e^-\to W^+ W^- b\bar b$ process  is obtained from the
$W^+bW^-\bar{b}$ cuts of the $e^+e^-$ forward-scattering amplitude. 
In the energy region $\sqrt{s}\approx 2 m_t$ 
 the amplitude is 
dominated by the production of resonant top quarks with small 
virtuality. This allows us to integrate out hard modes ($\sim m_t$)
 and represent the forward-scattering amplitude as the 
sum of two terms~\cite{Beneke:2003xh}, 
\begin{eqnarray}
\label{eq:master}
i {\cal A} &=&\sum_{k,l} C^{(k)}_p  C^{(l)}_p \int d^4 x \,
\braket{e^- e^+ |
\Tprod{i {\cal O}_p^{(k)\dagger}(0)\,i{\cal O}_p^{(l)}(x)}|e^- e^+} + \,\sum_{k} \,C_{4 e}^{(k)} 
\braket{e^- e^+|i {\cal O}_{4e}^{(k)}(0)|e^- e^+}. \;\;\;\;\;\;
\vspace*{-0.cm}
\end{eqnarray}

The matrix elements in (\ref{eq:master}) are evaluated in the 
``low-energy'' effective theory, which includes elements of 
soft-collinear and non-relativistic effective theory. The first term 
on the right-hand side of (\ref{eq:master}) describes the production 
of a resonant $t\bar t$ pair in terms of production (decay)
operators ${\cal O}_p^{(l)}(x)$ (${\cal O}_p^{(k)\dagger}(x)$) 
with short-distance coefficients $C^{(k,l)}_p$. The
second term accounts for the remaining non-resonant contributions,
which in the effective theory are described by four-electron production-decay
operators ${\cal O}^{(k)}_{4e}$. The calculation of 
the short-distance coefficients
$C^{(k)}_{4e}$ is performed in standard fixed-order perturbation
theory in the full electroweak theory.
 In particular, the 
top propagator is the free one not including the top width, 
which ensures that the amplitude depends only on the short-distance 
scales.
The leading 
imaginary parts of $C^{(k)}_{4e}$ arise from the cut two-loop
diagrams of order $\alpha_{\rm EW}^3$ shown in Fig.~\ref{fig1}. The corresponding contribution to the cross section is
\vskip -0.1cm
\begin{equation}
\label{eq:nonres}
\sigma_{\rm{non-res}} = \frac{1}{s}\,
\sum_k \,\mbox{Im}\left[C_{4 e}^{(k)}\right]\,
\braket{e^- e^+|i {\cal O}_{4e}^{(k)}(0)|e^- e^+}.
\end{equation}
Technically, this simply amounts to the calculation of the 
spin-averaged tree-level processes $e^+ e^- \to t W^- \bar b$ and 
$e^+ e^- \to \bar t W^+ b$ with no width supplied to the 
intermediate top-quark propagators. Instead, the divergence 
from the top-quark propagators going on-shell is regularized
dimensionally. Details on the computation and integral representations
of the result for (\ref{eq:nonres}) can be found in~\cite{Beneke:2010mp}.

Through the computation of the four-electron matching coefficients loose cuts ($\sim m_t$) on the
$bW^+$ and $\bar{b}W^-$ invariant masses can be incorporated easily, as it has
been discussed in the context of $W$-pair production near threshold~\cite{Actis:2008rb}.
The result  obtained in~\cite{Beneke:2010mp} covers the case of symmetric cuts on the invariant mass of the $bW$
subsystems  ($p^2_{bW}$)  of the form $m_t -\Delta M_t   \le \sqrt{p^2_{bW}} \le m_t +\Delta M_t$,
for $\Delta M_t\gg \Gamma_t$, up to the total cross section ($\Delta M_{t,\rm max}=m_t-M_W$).
An alternative approach has been developed in parallel~\cite{Hoang:2008ud,Hoang:2010gu} that includes
the effects of invariant-mass cuts on the $Wb$ pairs entirely through calculations in NRQCD. This works if the invariant-mass cuts around $m_t$ are neither very loose nor very tight, and provided 
that the non-resonant background processes are small (which at NLO was checked~\cite{Hoang:2010gu} by computing the
full  $e^+e^- \to W^+W^- b \bar{b}$ cross section at tree-level with MadGraph).
Under these assumptions, part of the $\alpha_s$-corrections to the non-resonant contributions has already been analyzed
in~\cite{Hoang:2010gu}, which in our approach correspond to NNLO contributions.
\vspace*{-.2cm}

\section{Results}
%
\begin{figure}[t]
\vskip -.5cm
\hskip -.1cm
\includegraphics[width=0.5\textwidth]{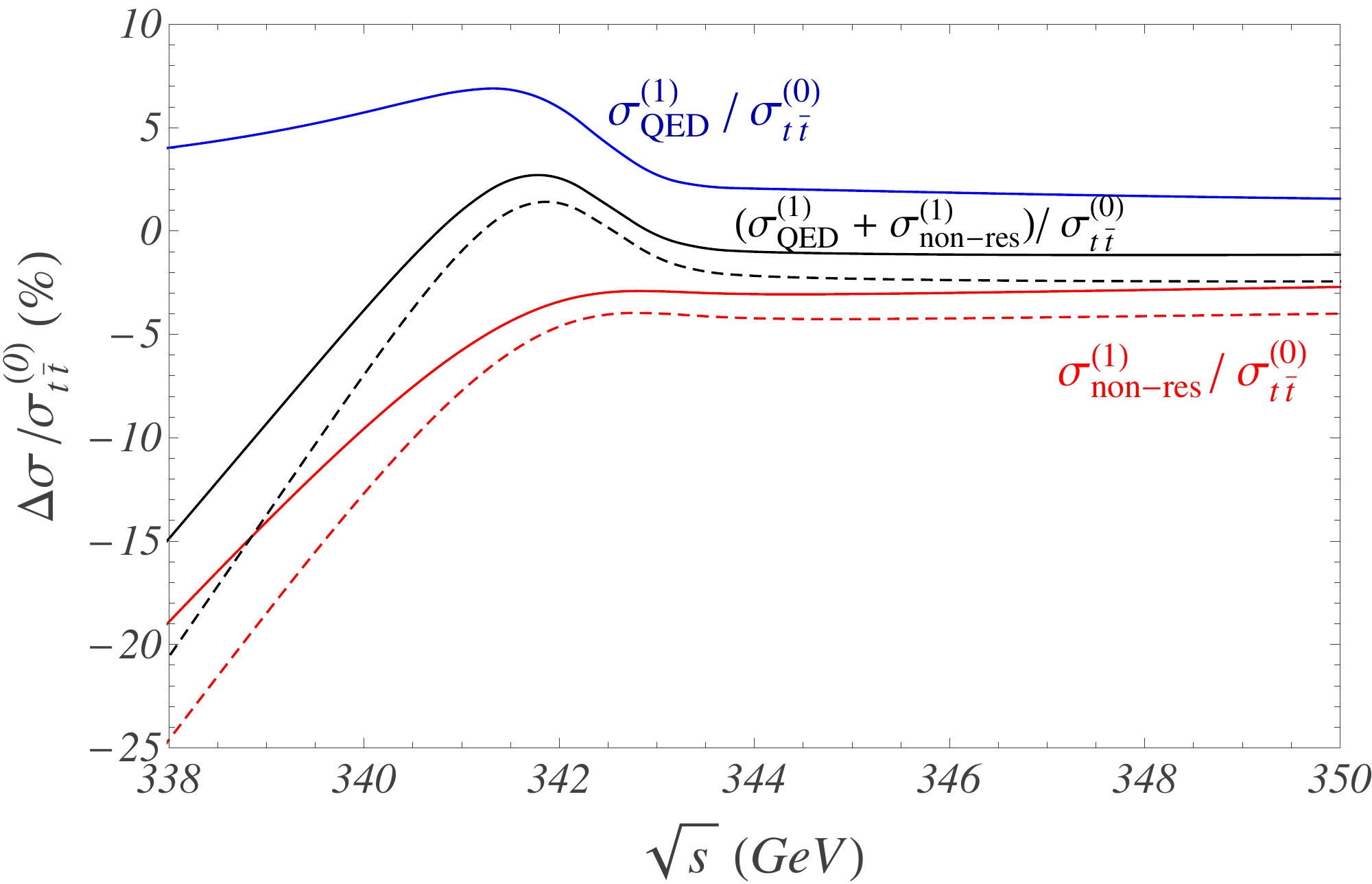}\\
{\vskip -5.38cm
\hskip 7.5cm
\includegraphics[width=0.482\textwidth]{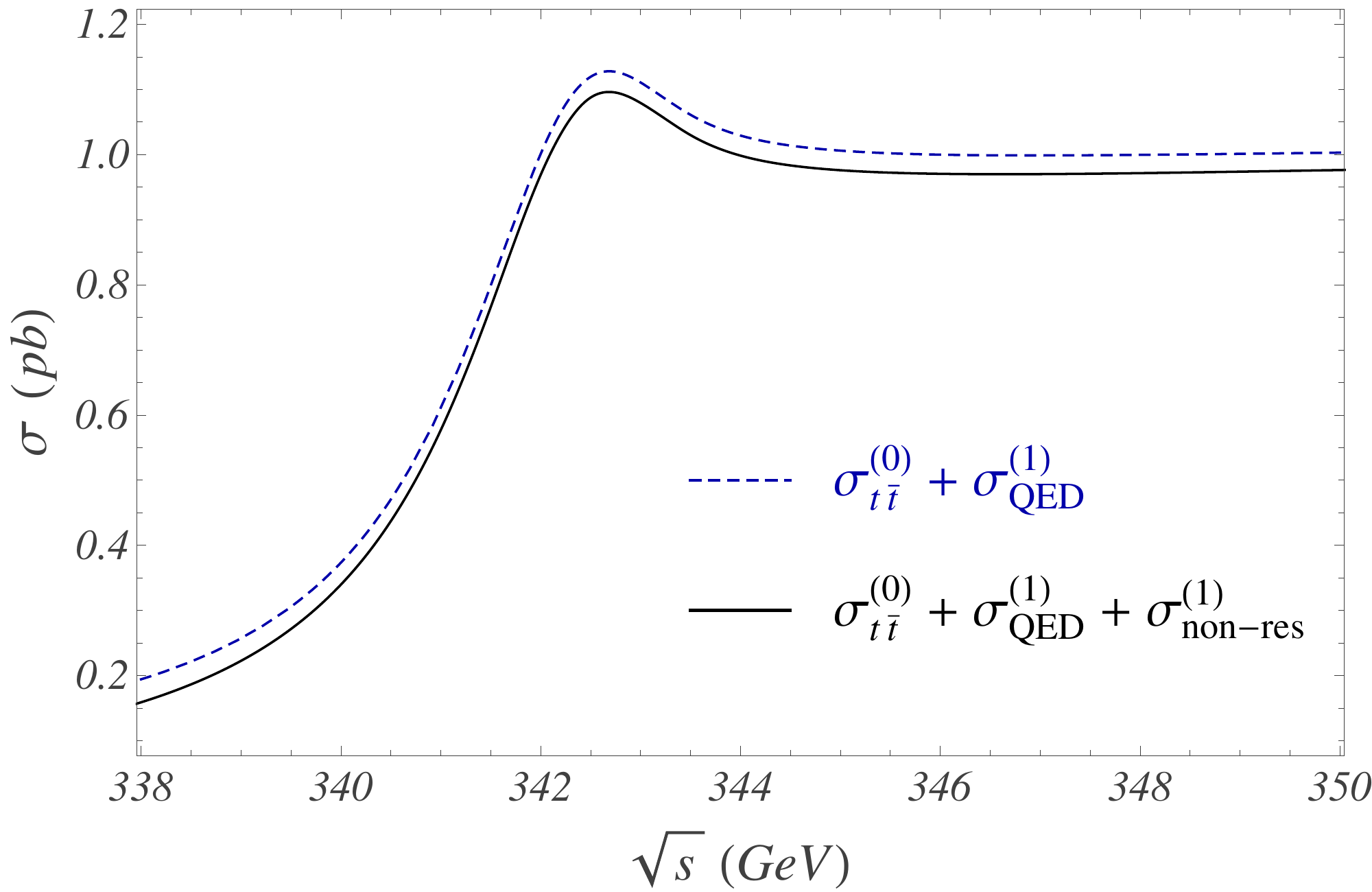}}
\vskip -.2cm
\caption{Left: Relative sizes of the QED, $\sigma^{(1)}_{\rm QED}$,
 and non-resonant, $\sigma^{(1)}_{\rm{non-res}}$, corrections
with  respect to the LO  cross 
section, $\sigma^{(0)}_{t\bar t}$, in percent, for $\Delta M_{t,\rm max}$ (solid) and $\Delta M_t=15$~GeV (dashed).
Right: Total cross section with LO QCD effects (dashed) and including NLO electroweak
corrections (solid) at energies close to threshold.
Input parameters: $m_t=172$~GeV, $\Gamma_t=1.47$~GeV and $\alpha_s(30~\rm{GeV})=0.142$.}
\label{fig2}
\vskip -.cm
\end{figure}
\vskip -.2cm
The left plot in Fig.~\ref{fig2} displays the relative sizes of the NLO electroweak
corrections with respect 
to the LO result 
for the $e^+e^-\to W^+W^-b\bar{b}$ cross section, which includes 
the summation of Coulomb corrections. 
The QED contribution represents a correction of about 2\% above threshold 
and rises to a maximum of 7\% just below the peak, while
the non-resonant contributions
give a constant negative shift of about 3\% above threshold. Below 
threshold the relative size of the non-resonant corrections is very large, 
since the LO result rapidly vanishes, reaching up to 19\%. Hence below 
threshold they represent the leading electroweak correction
to the total $t\bar{t}$ cross section. 
We observe a partial cancellation of the QED and
non-resonant corrections in the peak region and at energies above.
A sensitivity to the invariant-mass cut $\Delta M_t$ in the $bW^+$ and 
$\bar{b}W^-$ subsystem enters first at NLO through the non-resonant 
contributions. Restricting the available phase-space for the final-state 
particles by tightening the invariant-mass cuts $\Delta M_t$
makes the non-resonant contributions even more important. This is shown by 
the dashed lines in Fig.~\ref{fig2}, corresponding to 
$\Delta M_t=15$~GeV.

Aside from the pure QCD corrections, the NLO prediction 
for the $e^+e^-\to W^+W^-b\bar{b}$ total cross section is displayed by the 
solid line in the right plot of Fig.~\ref{fig2}. The absolute
size of the non-resonant correction is given by the difference between the 
dashed line, which only includes the QED NLO correction, and the solid one. 
This negative shift amounts to 27--35~fb  for $\sqrt{s}$ in the interval $(338,350)$~GeV.


\vspace*{-.2cm}
\begin{thebibliography}{99}

\bibitem{Martinez:2002st}
M.~Martinez and R.~Miquel,
\newblock Eur. Phys. J. {\bf C27}, 49 (2003).

\bibitem{Beneke:2008ec}
M.~Beneke, Y.~Kiyo and K.~Schuller,
PoS {\bf RADCOR2007}, 051 (2007).

\bibitem{Hoang:2001mm}
A.~H. Hoang, A.~V. Manohar, I.~W. Stewart and T.~Teubner,
\newblock Phys. Rev. {\bf D65}, 014014 (2002);\\
A.~Pineda and A.~Signer,
\newblock Nucl. Phys. {\bf B762}, 67 (2007).

\bibitem{Hoang:2004tg}
 A.~H.~Hoang and C.~J.~Reisser,
 Phys.\ Rev.\  D {\bf 71}, 074022 (2005). 

\bibitem{Beneke:2003xh}
M.~Beneke, A.~P. Chapovsky, A.~Signer and G.~Zanderighi,
\newblock Phys. Rev. Lett. {\bf 93}, 011602 (2004); 
M.~Beneke, A.~P. Chapovsky, A.~Signer and G.~Zanderighi,
\newblock Nucl. Phys. {\bf B686}, 205 (2004);\\
 M.~Beneke, P.~Falgari, C.~Schwinn, A.~Signer and G.~Zanderighi,
 Nucl.\ Phys.\  B {\bf 792}, 89 (2008).


\bibitem{Beneke:2010mp}
M.~Beneke, B.~Jantzen and P.~Ruiz-Femenia,
Nucl.\ Phys.\  B {\bf 840}, 186  (2010).

\bibitem{Actis:2008rb}
S.~Actis, M.~Beneke, P.~Falgari and C.~Schwinn,
Nucl.\ Phys.\  B {\bf 807}, 1 (2009).


\bibitem{Hoang:2008ud}
A.~H.~Hoang, C.~J.~Reisser and P.~Ruiz-Femenia,
Nucl.\ Phys.\ Proc.\ Suppl.\  {\bf 186}, 403 (2009).

\bibitem{Hoang:2010gu}
A.~H.~Hoang, C.~J.~Reisser and P.~Ruiz-Femenia,
Phys.\ Rev.\  D {\bf 82},  014005 (2010).

\end{thebibliography}
\end{document}